\documentclass{article}
\usepackage{graphicx}
\usepackage{amsmath}
\usepackage{amsfonts}
\usepackage{amssymb}
\input epsf

\begin{document}

\title{Modal expansions and non-perturbative quantum field theory in Minkowski space}
\author{Nathan Salwen\thanks{email: salwen@physics.harvard.edu}\\\textit{Harvard University}\\\textit{Cambridge, MA 02138}
\and Dean Lee\thanks{email: dlee@physics.umass.edu}\\\textit{University of Massachusetts}\\\textit{Amherst, MA 01003}}
\maketitle
\begin{abstract}
We introduce a spectral approach to non-perturbative field theory within the
periodic field formalism. As an example we calculate the real and imaginary
parts of the propagator in $1+1$ dimensional $\phi^{4}$ theory, identifying
both one-particle and multi-particle contributions. \ We discuss the
computational limits of existing diagonalization algorithms and suggest new
quasi-sparse eigenvector methods to handle very large Fock spaces and higher
dimensional field theories. \ [PACS numbers: \ 11.10Kk, 12.38Lg]
\end{abstract}

\section{Introduction}

Modal expansion methods have recently been used to study non-perturbative
phenomena in quantum field theory \cite{spher}-\cite{periodic}. Modal field
theory, the name for the\ general procedure, consists of two main parts. The
first is to approximate field theory as a finite-dimensional quantum
mechanical system. The second is to analyze the properties of the reduced
system using one of several computational techniques. \ The quantum mechanical
approximation is generated by decomposing field configurations into free wave
modes. \ This technique has been explored using both spherical partial waves
(spherical field theory \cite{spher}\cite{thirring}) and periodic box modes
(periodic field theory \cite{periodic}).

Having reduced field theory to a more tractable quantum mechanical system, we
have several different ways to proceed. \ Boson interactions in Euclidean
space, for example, can be modeled using the method of diffusion Monte Carlo.
\ In many situations, however, Monte Carlo techniques are inadequate. \ These
include unquenched fermion systems, processes in Minkowski space, and the
phenomenology of multi-particle states. \ Difficulties arise when the
functional integral measure cannot be treated as a probability distribution or
when information must be extracted from excited states obscured by dominant
lower lying states. \ Fortunately there are several alternative methods in the
modal field formalism which avoid these problems. \ Matrix Runge-Kutta
techniques were introduced in \cite{thirring} as a method for calculating
unquenched fermion interactions. \ Here we discuss a different approach, one
which directly calculates the spectrum and eigenstates of the Hamiltonian.
\ For this approach it is essential that the Hamiltonian is time-independent,
and so we will consider periodic rather than spherical field theory. \ As we
demonstrate, these methods naturally accommodate the study of multi-particle
states and Minkowskian dynamics.

We apply the spectral approach to $1+1$ dimensional $\phi^{4}$ theory in a
periodic box and calculate the real and imaginary parts of the $\phi$
propagator. \ Some interesting properties of $\phi_{2}^{4}$ theory such as the
phase transition at large coupling were already discussed within the modal
field formalism using Euclidean Monte Carlo techniques \cite{periodic}. \ The
purpose of this analysis is of a more general and exploratory nature. \ Our
aim is to test the viability of modal diagonalization techniques for quantum
field Hamiltonians. \ We would like to know whether we can clearly see
multi-particle phenomena, the size of the errors and computational limitations
with current computer resources, and how such methods might be extended to
more complicated higher dimensional field theories.

The spectral method presented in the first part of our analysis is similar to
the work of Brooks and Frautschi \cite{brooks},\footnote{We thank the referee
of the original manuscript for providing information on this reference.} who
considered a $1+1$ dimensional Yukawa model in a periodic box and deserves
credit for the first application of diagonalization techniques using plane
wave modes in a periodic box. \ Our calculations are also similar in spirit to
diagonalization-based Hamiltonian lattice formulations \cite{hollenberg} and
Tamm-Dancoff light-cone and discrete\ light-cone quantization \cite{perry}%
-\cite{lightcone}. \ There are, however, some differences which we should
mention. \ As in \cite{brooks} we are using a momentum lattice rather than a
spatial lattice. \ We find this convenient to separate out invariant subspaces
according to total momentum quantum numbers. \ Since we are using an equal
time formulation our eigenvectors are not boost invariant as they would be on
the light cone. \ Also we are using a simple momentum cutoff scheme rather
than a regularization scheme which includes Tamm-Dancoff Fock-space
truncation. \ As a result our renormalization procedure is relatively
straightforward, but we will have to confront the problem of diagonalizing
large Fock spaces from the very beginning. \ In the latter part of the paper
we mention current work on quasi-sparse eigenvector methods which can handle
even exceptionally large Fock spaces. \ Despite the differences among the
various diagonalization approaches to field theory, the issues and problems
discussed in our analysis are of a general nature. \ We hope that the ideas
presented here will be of use for the various different approaches.

\section{Spectral method}

The field configuration $\phi$ in $1+1$ dimensions is subject to periodic
boundary conditions $\phi(t,x-L)=\phi(t,x+L).$ \ Expanding in terms of
periodic box modes, we have
\begin{equation}
\phi(t,x)=\sqrt{\tfrac{1}{2L}}%
%TCIMACRO{\dsum \limits_{n=0,\pm1,...}}%
%BeginExpansion
{\displaystyle\sum\limits_{n=0,\pm1,...}}
%EndExpansion
\phi_{n}(t)e^{in\pi x/L}.
\end{equation}
The sum over momentum modes is regulated by choosing some large positive
number $N_{\max}$ and throwing out all high-momentum modes $\phi_{n}$ such
that $\left|  n\right|  >N_{\max}$. In this theory renormalization can be
implemented by normal ordering the $\phi^{4}$ interaction term. After a
straightforward calculation (details are given in \cite{periodic}), we find
that the counterterm Hamiltonian has the form
\begin{equation}
\tfrac{6\lambda b}{4!2L}%
%TCIMACRO{\dsum \limits_{n=-N_{\max},N_{\max}}}%
%BeginExpansion
{\displaystyle\sum\limits_{n=-N_{\max},N_{\max}}}
%EndExpansion
\phi_{-n}\phi_{n},
\end{equation}
where
\begin{equation}
b=%
%TCIMACRO{\dsum \limits_{n=-N_{\max},N_{\max}}}%
%BeginExpansion
{\displaystyle\sum\limits_{n=-N_{\max},N_{\max}}}
%EndExpansion
\tfrac{1}{2\omega_{n}},\qquad\omega_{n}=\sqrt{\tfrac{n^{2}\pi^{2}}{L^{2}}%
+\mu^{2}}.
\end{equation}
We represent the canonical conjugate pair $\phi_{n}$ and $\frac{d\phi_{-n}%
}{dt}$ using the Schr\"{o}dinger operators $q_{n}$ and $-i\tfrac{\partial
}{\partial q_{n}}$. \ Then the functional integral for $\phi^{4}$ theory is
equivalent to that for a quantum mechanical system with Hamiltonian
\begin{align}
H  &  =%
%TCIMACRO{\dsum \limits_{n=-N_{\max},N_{\max}}}%
%BeginExpansion
{\displaystyle\sum\limits_{n=-N_{\max},N_{\max}}}
%EndExpansion
\left[  -\tfrac{1}{2}\tfrac{\partial}{\partial q_{-n}}\tfrac{\partial
}{\partial q_{n}}+\tfrac{1}{2}(\omega_{n}^{2}-\tfrac{\lambda b}{4L}%
)q_{-n}q_{n}\right] \\
&  +\tfrac{\lambda}{4!2L}%
%TCIMACRO{\dsum \limits_{\genfrac{}{}{0pt}{1}{n_{i}=-N_{\max},N_{\max}%
%}{n_{1}+n_{2}+n_{3}+n_{4}=0}}}%
%BeginExpansion
{\displaystyle\sum\limits_{\genfrac{}{}{0pt}{1}{n_{i}=-N_{\max},N_{\max
}}{n_{1}+n_{2}+n_{3}+n_{4}=0}}}
%EndExpansion
q_{n_{1}}q_{n_{2}}q_{n_{3}}q_{n_{4}}.\nonumber
\end{align}

We now consider the Hilbert space of our quantum mechanical system. \ Given
$d$, an array of non-negative integers,%
\begin{equation}
d=\left\{  d_{-N_{\max}},\cdots d_{N_{\max}}\right\}  ,
\end{equation}
we denote $p_{d}(q)$ as the following monomial with total degree $\left|
d\right|  $,%

\begin{equation}
p_{d}(q)=%
%TCIMACRO{\dprod \limits_{n=-N_{\max},N_{\max}}}%
%BeginExpansion
{\displaystyle\prod\limits_{n=-N_{\max},N_{\max}}}
%EndExpansion
q_{n}^{d_{n}},\qquad%
%TCIMACRO{\dsum \limits_{n}}%
%BeginExpansion
{\displaystyle\sum\limits_{n}}
%EndExpansion
d_{n}=\left|  d\right|  .
\end{equation}
We define $G_{\zeta}(q)$ to be a Gaussian of the form\footnote{$G_{\zeta}(q)$
has been defined such that $G_{\mu}(q)$ is the ground state of the free
theory.}%
\begin{equation}
G_{\zeta}(q)=%
%TCIMACRO{\dprod \limits_{n=-N_{\max},N_{\max}}}%
%BeginExpansion
{\displaystyle\prod\limits_{n=-N_{\max},N_{\max}}}
%EndExpansion
\exp\left[  -\tfrac{q_{-n}q_{n}\sqrt{\zeta^{2}+n^{2}\pi^{2}/L^{2}}}{2}\right]
.
\end{equation}
$\zeta$ is an adjustable parameter which we will set later. \ Any
square-integrable function $\psi(q)$ can be written as a superposition%

\begin{equation}
\psi(q)=%
%TCIMACRO{\dsum \limits_{d}}%
%BeginExpansion
{\displaystyle\sum\limits_{d}}
%EndExpansion
c_{d}\,p_{d}(q)G_{\zeta}(q). \label{sum}%
\end{equation}
In this analysis we consider only the zero-momentum subspace. \ We impose this
constraint by restricting the sum in (\ref{sum}) to monomials satisfying
\begin{equation}%
%TCIMACRO{\dsum \limits_{n}}%
%BeginExpansion
{\displaystyle\sum\limits_{n}}
%EndExpansion
nd_{n}=0.
\end{equation}
\ 

We will restrict the space of functions $\psi(q)$ further by removing high
energy states in the following manner. \ Let
\begin{equation}
k(d)=%
%TCIMACRO{\dsum _{n}}%
%BeginExpansion
{\displaystyle\sum_{n}}
%EndExpansion
\left|  n\right|  d_{n}.
\end{equation}
$k(d)$ was first introduced in \cite{thirring} and provides an estimate of the
kinetic energy associated with a given state. Let us define two auxiliary
cutoff parameters, $K_{\max}$ and $D_{\max}$. \ We restrict the sum in
(\ref{sum}) to monomials such that $k(d)<K_{\max}$ and $\left|  d\right|  \leq
D_{\max}$. \ We will refer to the corresponding subspace as $V_{K_{\max
},D_{\max}}$. \ The cutoff $K_{\max}$ removes states with high kinetic energy
and the cutoff $D_{\max}$ eliminates states with a large number of excited
modes.\footnote{In the case of spontaneous symmetry breaking, the broken
symmetry of the vacuum may require retaining a large number of $q_{0}$ modes.
\ This however is remedied by shifting the variable, $q_{0}^{\prime}%
=q_{0}-\left\langle 0\right|  q_{0}\left|  0\right\rangle $.} \ We should
stress that $K_{\max}$ and $D_{\max}$ are only auxiliary cutoffs. \ We
increase these parameters until the physical results appear close to the
asymptotic limit $K_{\max},D_{\max}\rightarrow\infty$. \ In our scheme
ultraviolet regularization is provided only by the momentum cutoff parameter
$N_{\max}$.

Our plan is to analyze the spectrum and eigenstates of $H$ restricted to this
approximate low energy subspace, $V_{K_{\max},D_{\max}}$. For any fixed $L$
and $N_{\max}$, $H$ is the Hamiltonian for a finite-dimensional quantum
mechanical system and the results should converge in the limit $K_{\max
},D_{\max}\rightarrow\infty$. \ We obtain the desired field theory result by
then taking the limit $L,\frac{N_{\max}}{L}\rightarrow\infty$.

\section{Results}

We have calculated the matrix elements of $H$ restricted to $V_{K_{\max
},D_{\max}}$ using a symbolic differentiation-integration
algorithm\footnote{All codes can be obtained by request from the authors.} and
diagonalized $H$, obtaining both eigenvalues and eigenstates. \ Let $\Delta$
be the full propagator,
\begin{equation}
\Delta(p^{2})=i\int d^{2}x\,e^{ip_{\nu}x^{\nu}}\left\langle 0\right|  T\left[
\phi(x^{\mu})\phi(0)\right]  \left|  0\right\rangle .
\end{equation}
We have computed $\Delta$ by inserting our complete set of eigenstates
(complete in $V_{K_{\max},D_{\max}}$)$.$ \ Let $\Delta_{\text{mp}}$ be the
multi-particle contribution to $\Delta$,%
\begin{equation}
\Delta_{\text{mp}}(p^{2})=\Delta(p^{2})-\Delta_{\text{pole}}(p^{2}),
\end{equation}
where $\Delta_{\text{pole}}$ is the single-particle pole contribution. \ We
are primarily interested in $\Delta_{\text{mp}}$, a quantity that cannot be
obtained for $p^{2}>0$ using Monte Carlo methods. \ Since the imaginary part
of $\Delta_{\text{pole}}$ is a delta function, it is easy to distinguish the
single-particle and multi-particle contributions in a plot of the imaginary
part of $\Delta$. \ The real part of $\Delta$, however, is dominated by the
one-particle pole. \ For this reason we have chosen to plot the real part of
$\Delta_{\text{mp}}$ rather than that of $\Delta$.

Although we have referred to multi-particle states, it should be noted that in
our finite periodic box there are no true continuum multi-particle states.
\ Instead we find densely packed discrete spectra with level separation of
size $\sim L^{-1}$ which become continuum states in the limit $L\rightarrow
\infty$. \ We can approximate the contribution of these $L\rightarrow\infty$
continuum states by a simple smoothing process. We included a small width
$\Gamma\sim L^{-1}$ to each of the would-be continuum states and averaged over
a range of values for $L$. \ For the results we report here we have averaged
over values $L=2.0\pi,2.1\pi,\cdots2.8\pi.$ \ For convenience all units have
been scaled such that $\mu=1$.

The parameter $\zeta$ was adjusted to reduce the errors due to the finite
cutoff values $K_{\max}$ and $D_{\max}$. \ Since the spectrum of $H$ is
bounded below, errors due to finite $K_{\max}$ and $D_{\max}$ generally drive
the estimated eigenvalues higher. \ One strategy, therefore, is to optimize
$\zeta$ by minimizing the trace of $H$ restricted to the subspace $V_{K_{\max
},D_{\max}}$. \ The approach used here is a slight variation of this --- we
have minimized the trace of $H$ restricted to a smaller subspace $V_{K_{\max
}^{\prime},D_{\max}^{\prime}}\subset V_{K_{\max},D_{\max}}$. \ The aim is to
accelerate the convergence of the lowest energy states rather than the entire
space $V_{K_{\max},D_{\max}}$. \ Throughout our analysis we used $K_{\max
}^{\prime},D_{\max}^{\prime}=8,3.$

For $\frac{\lambda}{4!}=0.50$ we have plotted the imaginary part of $\Delta$
in Figure 1 and the real part of $\Delta_{\text{mp}}$ in Figure 2$.$ \ The
value $\frac{\lambda}{4!}=0.50$ is above the threshold for reliable
perturbative approximation\footnote{For momenta $\left|  p^{2}\right|
\gtrsim1$.} ($\frac{\lambda}{4!}\lesssim0.25$) but below the critical value at
which $\phi\rightarrow-\phi$ symmetry breaks spontaneously ($\frac{\lambda
}{4!}\approx2.5).$ \ The contribution of the one-particle state appears near
$p^{2}=(0.93)^{2}$ and the three-particle threshold is at approximately
$p^{2}=(2.9)^{2}$. \ We have chosen several different values for $N_{\max
},K_{\max},D_{\max}$ to show the convergence as these parameters become large.
\ The plot for $N_{\max},K_{\max},D_{\max}=9,19,7$ appears relatively close to
the asymptotic limit. \ The somewhat bumpy texture of the curves is due to the
finite size of our periodic box and diminishes with increasing $L$. \ From
dimensional power counting, we expect errors for finite $N_{\max}$ to scale as
$N_{\max}^{-2}$. \ Assuming that $K_{\max}$ and $D_{\max}$ also function as
uniform energy cutoffs, we expect a similar error dependence -- and it appears
plausible from the results in Figures 1 and 2. \ A more systematic analysis of
the errors and extrapolation methods for finite $N_{\max},K_{\max},D_{\max}$,
and $L,$ will be discussed in future work.

In Figures 3 and 4 we have compared our spectral calculations with the
two-loop perturbative result for $\frac{\lambda}{4!}=0.01$. \ We have used
$N_{\max},K_{\max},D_{\max}=9,19,7$, and the agreement appears good. \ For
small $\frac{\lambda}{4!}$ the propagator has a very prominent logarithmic
cusp at the three-particle threshold, which can be seen clearly in Figures 3
and 4.

In Figures 5 and 6 we have compared results for $\frac{\lambda}{4!}=0.25$,
$0.50$, $1.00$. \ We have again used $N_{\max},K_{\max},D_{\max}=9,19,7$. \ In
contrast with the quadratic scaling in the perturbative regime, the results
here scale approximately linearly with $\frac{\lambda}{4!}$. \ An interesting
and perhaps related observation is that the magnitude of the multi-particle
contribution to $\Delta$ remains small $(\lesssim0.003$) even for the rather
large coupling value $\frac{\lambda}{4!}=1.00$.

\section{Limitations and new ideas}

We now address the computational limits of the diagonalization techniques
presented in this work. \ These techniques are rather straightforward and can
in principle be generalized to any field theory. \ In practise however the
Fock space $V_{K_{\max},D_{\max}}$ becomes prohibitively large, especially for
higher dimensional theories. \ The data in Figures 1 and 2 and crosschecks
with Euclidean Monte Carlo results\footnote{See \cite{periodic} for a
discussion of these methods.} suggest that for $N_{\max}=9\ $and
$L=2.0\pi,\cdots2.8\pi$ our spectral results with $K_{\max},D_{\max}=19,7$ and
$\frac{\lambda}{4!}<1$ are within 20\% of the $K_{\max},D_{\max}%
\rightarrow\infty$ limit. $\ $In this case $V_{K_{\max},D_{\max}}$ is a $2036$
dimensional space and requires about 100 MB of RAM using general (dense)
matrix methods.

Sparse matrix techniques such as the Lanczos or Arnoldi schemes allow us to
push the dimension of the Fock space to about 10$^{5}$ states. \ This may be
sufficient to do accurate calculations near the critical point $\frac{\lambda
}{4!}\approx2.5$ for larger values of $L$ and $N_{\max}.$ \ It is, however,
near the upper limit of what is possible using current computer technology and
existing algorithms. \ Unfortunately field theories in $2+1$ and $3+1$
dimensions will require much larger Fock spaces, probably at least 10$^{12}$
and 10$^{18}$ states respectively. \ In order to tackle these larger Fock
spaces it is necessary to venture beyond standard diagonalization approaches.
\ The problem of large Fock spaces ($\gg$10$^{6}$ states) is beyond the
intended scope of this analysis. \ But since it is of central importance to
the diagonalization approach to field theory we would like to briefly comment
on current work being done which may resolve many of the difficulties. \ The
new approach takes advantage of the sparsity of the Fock-space Hamiltonian and
the approximate (quasi-)sparsity of the eigenvectors. \ A detailed description
will be provided in a future publication \cite{new}.

We start with some observations about the eigenvectors of the $\phi_{1+1}^{4}$
Hamiltonian for $N_{\max}=9,$ $L=2.5\pi$ and $K_{\max},D_{\max}=19,7.$ \ To
make certain that we are probing non-perturbative physics we will set
$\frac{\lambda}{4!}=2.5,$ the approximate critical point value. \ We label the
normalized eigenvectors as $\left|  v_{0}\right\rangle $, $\left|
v_{1}\right\rangle $,$\cdots,$ ascending in order with respect to energy. \ We
also define $\left|  b_{0}\right\rangle $, $\left|  b_{1}\right\rangle
$,$\cdots$ as the normalized eigenvectors of the free, non-interacting theory.
\ For any $v_{i}$ we know
\begin{equation}
\sum_{j}\left|  \left\langle b_{j}|v_{i}\right\rangle \right|  ^{2}=1.
\end{equation}
Let us define $\left\|  \left|  v_{i}\right\rangle \right\|  _{(n)}$ as the
partial sum%
\begin{equation}
\left\|  \left|  v_{i}\right\rangle \right\|  _{(n)}=\sum_{k=1,\cdots
n}\left|  \left\langle b_{j_{k}}|v_{i}\right\rangle \right|  ^{2},
\end{equation}
where the inner products have been sorted from largest to smallest
\begin{equation}
\left|  \left\langle b_{j_{1}}|v_{i}\right\rangle \right|  \geq\left|
\left\langle b_{j_{2}}|v_{i}\right\rangle \right|  \geq\cdots.
\end{equation}
Table 1 shows $\left\|  \left|  v_{i}\right\rangle \right\|  _{(n)}$ for
several eigenvectors and different values of $n$.
\[
\overset{\text{Table 1}}{%
\begin{array}
[c]{c}%
\\%
\begin{tabular}
[c]{l|l|l|l|l|}%
$\left\|  \left|  v_{i}\right\rangle \right\|  _{(n)}$ & $n=10$ & $n=20$ &
$n=40$ & $n=80$\\\hline
$\left|  v_{0}\right\rangle $ & 0.75 & 0.84 & 0.90 & 0.94\\\hline
$\left|  v_{1}\right\rangle $ & 0.89 & 0.92 & 0.95 & 0.97\\\hline
$\left|  v_{5}\right\rangle $ & 0.87 & 0.91 & 0.94 & 0.96\\\hline
$\left|  v_{10}\right\rangle $ & 0.77 & 0.86 & 0.90 & 0.94\\\hline
\end{tabular}
\end{array}
}%
\]
\ Despite the non-perturbative coupling and complex phenomena associated with
the phase transition, we see from the table that each of the eigenvectors can
be approximated by just a small number of its largest Fock-space components.
\ We recall that the Fock space for this system has 2036 dimensions. \ The
eigenvectors are therefore quasi-sparse in this space, a consequence of the
sparsity of the Hamiltonian. \ If we write the Hamiltonian as a matrix in the
free Fock-space basis, a typical row or column contains only about 200
non-zero entries, a number we refer to as $N_{\text{transition}}.$ \ In
\cite{new} we show that a typical eigenvector will be dominated by the largest
$\sqrt{N_{\text{transition}}}$ elements.\footnote{There are some special
exceptions to this rule and they are discussed in \cite{new}. \ But these are
typically not relevant for the lower energy eigenstates of a quantum field
Hamiltonian.} \ The key point is that $\sqrt{N_{\text{transition}}}$ is quite
manageable --- on the order of $10^{3}$ and 10$^{5}$ for $2+1$ and $3+1$
dimensional field theories respectively. \ Although the size of the Fock space
for these systems are enormous, the extreme sparsity of the Hamiltonian
suggests that the eigenvectors can be approximated using current computational resources.

With this simplification, the task is to find the important basis states for a
given eigenvector. \ Since the important basis states for one eigenvector are
generally different from that of another, each eigenvector is determined
independently. \ This provides a starting point for parallelization, and many
eigenvectors can be determined at the same time using massively parallel
computers. \ In \cite{new} we present a simple stochastic algorithm where the
exact eigenvectors act as stable fixed points of the update process.

\section{Summary}

We have introduced a spectral approach to periodic field theory and used it to
calculate the propagator in $1+1$ dimensional $\phi^{4}$ theory. \ We find
that the straightforward application of these methods with existing computer
technology can be useful for describing the multi-particle properties of the
theory, information difficult to obtain using Euclidean Monte Carlo methods.
\ However the extension to higher dimensional theories is made difficult by
the large size of the corresponding Fock space. \ As a possible solution to
this problem, we note that each eigenvector of the $\phi_{1+1}^{4}$
Hamiltonian can be well-approximated using relatively few components and
discuss some current work on quasi-sparse eigenvector
methods.{\normalsize \bigskip}

\noindent{\Large \textbf{Acknowledgment}}{\normalsize \bigskip}

\noindent We thank Eugene Golowich, Mark Windoloski, and Daniel Lee for useful
discussions. \ D.L. also thanks the organizers and participants of the CSSM
Workshop on Light-Cone QCD and Non-perturbative Hadron Physics in Adelaide.
\ We acknowledge financial support provided by the NSF under Grant 5-22968 and PHY-9802709.

\noindent{\Large \textbf{Figures}}\bigskip

\begin{itemize}
\item [Figure 1:]Imaginary part of $\Delta(p^{2})$ for $\frac{\lambda}%
{4!}=0.50$ and several values for $N_{\max},K_{\max},D_{\max}.$

\item[Figure 2:] Real part of $\Delta_{\text{mp}}(p^{2})$ for $\frac{\lambda
}{4!}=0.50$ and several values for $N_{\max},K_{\max},D_{\max}.$

\item[Figure 3:] Imaginary part of $\Delta(p^{2})$ for $\frac{\lambda}%
{4!}=0.01$ and comparison with the two-loop result$.$

\item[Figure 4:] Real part of $\Delta_{\text{mp}}(p^{2})$ for $\frac{\lambda
}{4!}=0.01$ and comparison with the two-loop result.

\item[Figure 5:] Imaginary part of $\Delta(p^{2})$ for $\frac{\lambda}%
{4!}=0.25,0.50,1.00$.

\item[Figure 6:] Real part of $\Delta_{\text{mp}}(p^{2})$ for $\frac{\lambda
}{4!}=0.25,0.50,1.00$.
\end{itemize}

\newpage

\begin{figure}[ptb]
\epsfbox{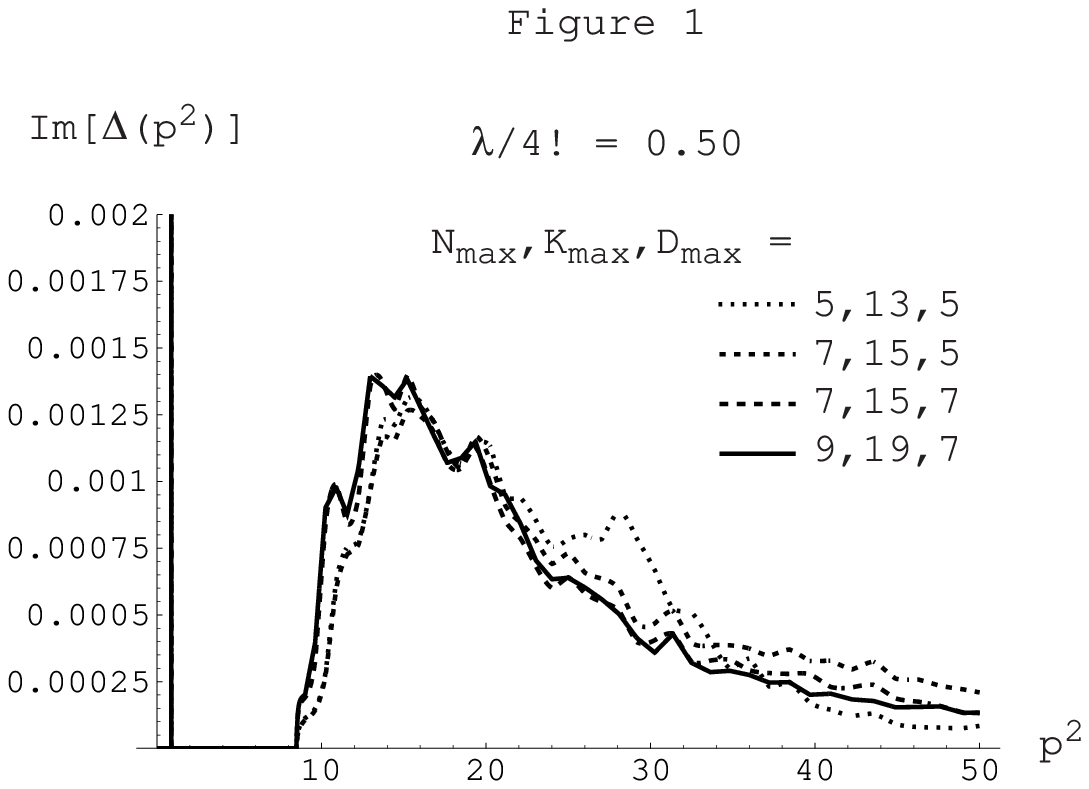}\end{figure}\begin{figure}[ptbptb]
\epsfbox{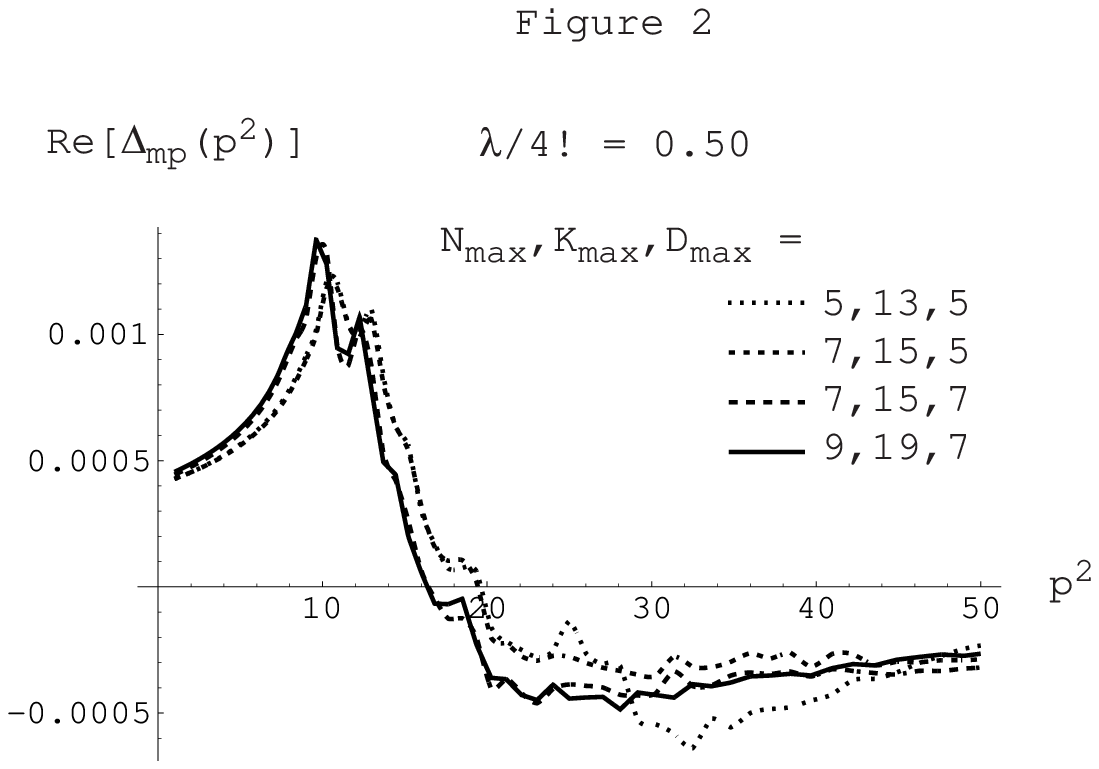}\end{figure}\begin{figure}[ptbptbptb]
\epsfbox{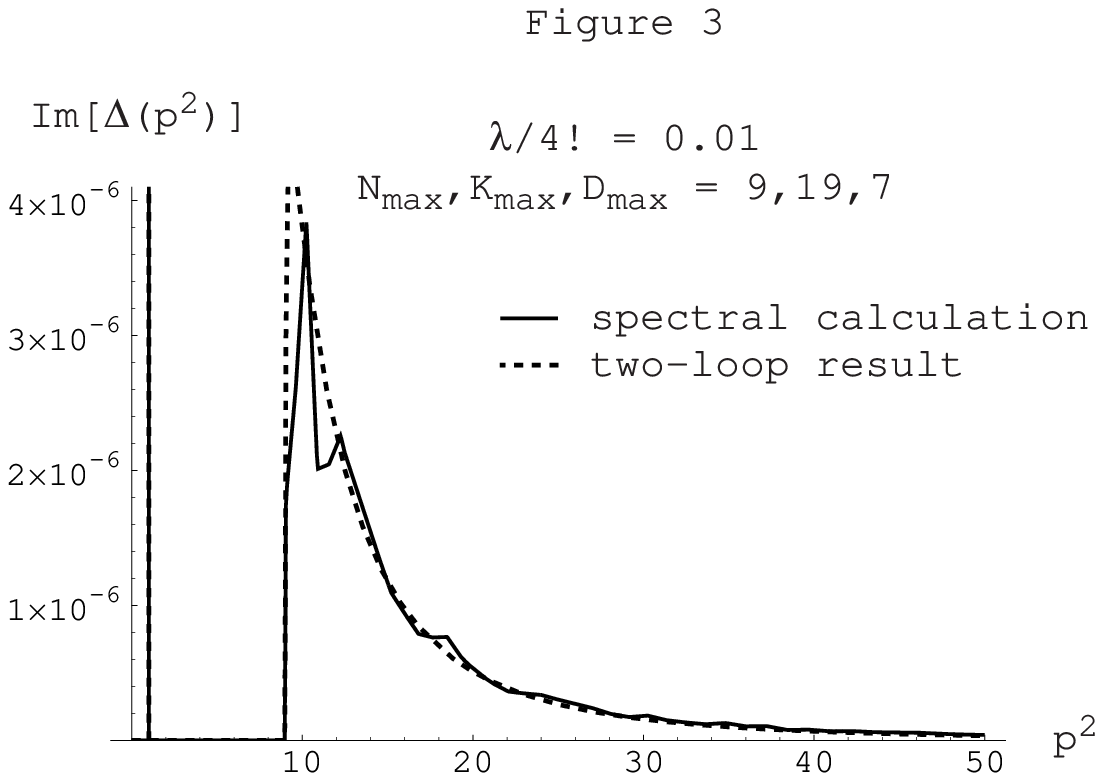}\end{figure}\begin{figure}[ptbptbptbptb]
\epsfbox{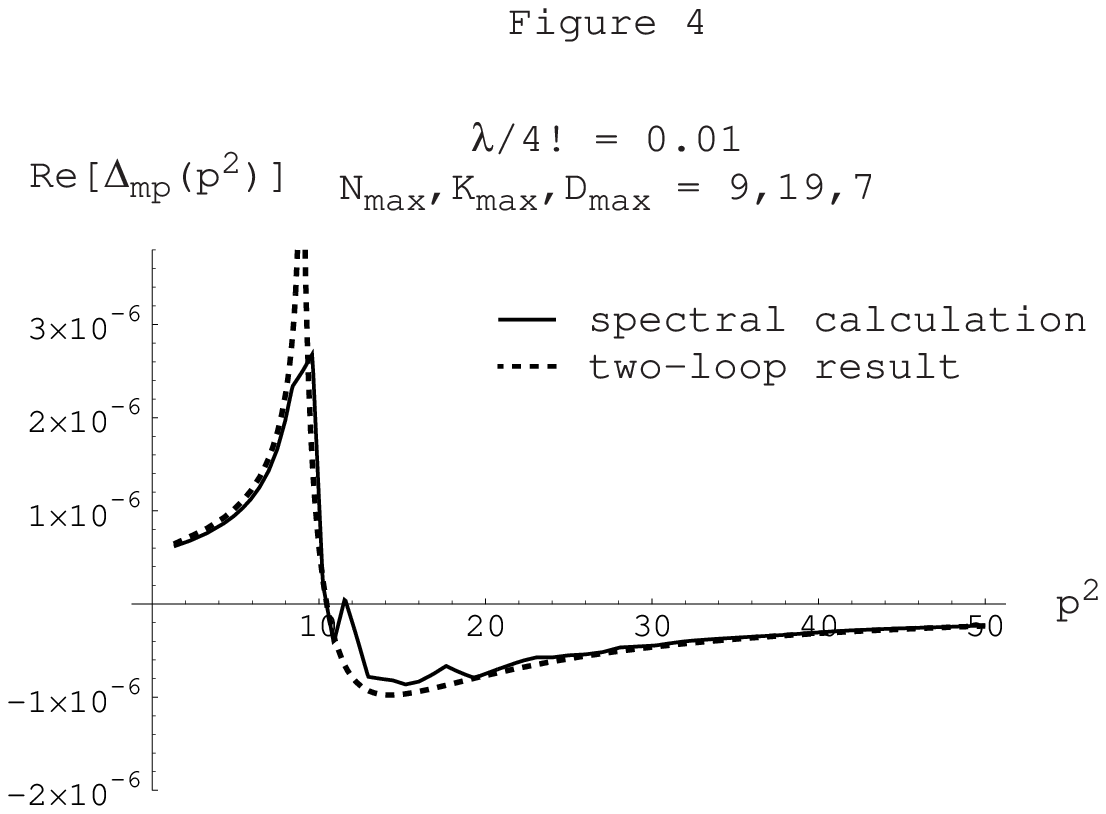}\end{figure}\begin{figure}[ptbptbptbptbptb]
\epsfbox{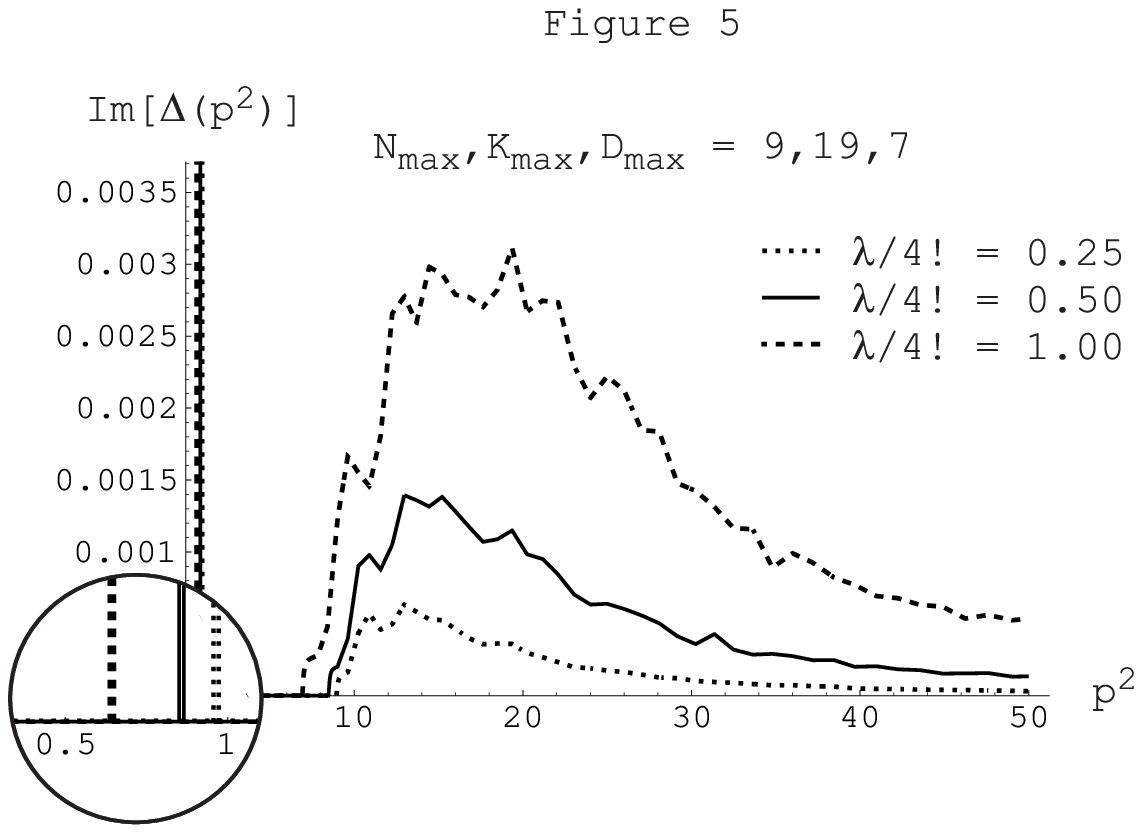}\end{figure}\begin{figure}[ptbptbptbptbptbptb]
\epsfbox{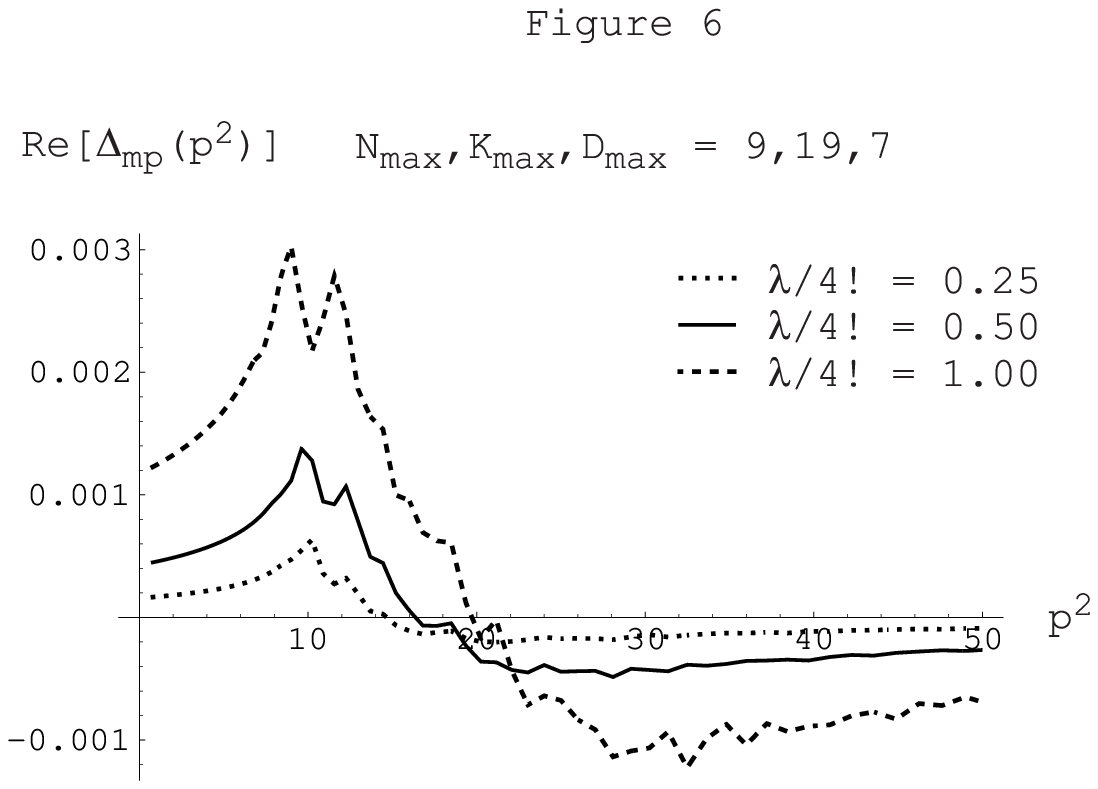}\end{figure}
\end{document}